# Landau-level spectrum and the effect of spin-orbit coupling in monolayer graphene on transition metal dichalcogenides

Qing Rao, Hongxia Xue*, and Dong-Keun Ki*

Department of Physics and HK Institute of Quantum Science & Technology, The University of Hong Kong, Pokfulam Road, Hong Kong, China
E-mail: hxxue@hku.hk, dkki@hku.hk

Keywords: graphene, quantum Hall effect, Landau level crossing, spin-orbit coupling, van der Waals heterostructures

**Abstract**

In graphene on transition metal dichalcogenides, two types of spin-orbit coupling (SOC)—Rashba and spin-valley Zeeman SOCs—can coexist that modify graphene's electronic band differently. Here, we show that the Landau levels (LLs) are also affected by these SOCs distinctively enough to estimate their relative strengths from the Landau fan diagram. Using a simple theoretical model, we calculated the LL spectrums of graphene for different SOC strengths, and found that when the total SOC is strong enough (i.e., when it is comparable to the half of the energy gap between the LLs of an intrinsic graphene), the corresponding LLs will split and cross with others depending sensitively on the relative strengths of the SOC terms. To demonstrate how one can use it to estimate the relative SOC strengths, we first identified the four key features that are well separated from the complex background and can be compared with experiment directly, and used them to show that in our sample, the Rashba SOC is stronger than the spin-valley Zeeman SOC that is consistent with other spectroscopic measurements. Our study therefore provides a simple and practical strategy to analyze the LL spectrum in graphene with SOC before carrying out more in-depth measurements.

## 1. Introduction

Interfacial interactions between the layers in van der Waals (vdW) heterostructures have been shown to be highly efficient in inducing new electronic properties that are originally absent in the individual layers. Examples include, but not limited to, strongly correlated states in magic-angle twisted bilayer graphene,[1,2] Hofstadter's butterfly effect in graphene/hexagonal boron-nitride (hBN) moiré superlattices,[3-5] and unconventional ferroelectricity at the interfaces between hBN layers[6-8] or between transition metal dichalcogenide (TMDC) layers.[9,10] Another system of interest is graphene-TMDC heterostructures (**Figure 1**), where the TMDC induces two different types of spin-orbit coupling (SOC) in graphene through proximity:[11-13] the spin-valley Zeeman term that couples out-of-plane spin and valley degrees of freedom, $H_{vz} = \lambda_{vz} \tau_z s_z$, and the Rashba term that couples in-plane spin and sublattice degrees of freedom, $H_R = \lambda_R \left( -\tau_z \sigma_x s_y - \sigma_y s_x \right)$.[12,14,15] Here, $\boldsymbol{\sigma} = (\sigma_x, \sigma_y, \sigma_z)$ and $\boldsymbol{s} = (s_x, s_y, s_z)$ are Pauli matrix vectors that act on the sublattice ($A$ and $B$) and spin (↑ and ↓) degrees of freedom in graphene, respectively, and $\tau_z = \pm 1$ represents $K$ and $K'$ valleys. Therefore, the graphene-TMDC heterostructures provide a new platform with high electron and hole mobility—in addition to the semiconductor-based 2D electron gas (2DEG)[16-19] systems—to investigate the effect of SOC on carrier transport in low-dimensional systems.

Many efforts have been made to estimate the SOC strengths of the system, which can be grouped into two: one that measures the spin relaxation time ($\tau_s$) of the charge carriers and uses a model to estimate the SOC strength from the measured $\tau_s$[11,20-22] and another that directly detects the change of the graphene's electronic band by the SOC.[13,23,24] These studies, however, often require a series of measurements by controlling more than two different experimental conditions such as a magnetic field (or its direction), charge density ($n$), and temperature ($T$). For instance, under a sufficiently large perpendicular magnetic field ($B$), the Landau levels (LLs) can be formed in graphene whose spectrum—the dispersion of the LLs in energy and magnetic fields—varies with the band structure and the internal degeneracy. In this quantum Hall (QH) regime, when the Fermi energy lies between the two LLs with an energy gap ($\Delta E_v$) that is larger than the thermal energy and the disorder width ($\Gamma$), $\Delta E_v > \max(kT, \Gamma)$ (**Figure 1(b)**; $k$: Boltzmann constant), the longitudinal resistance $R_{xx}$ shows a dip (i.e., a local minimum) at the corresponding filling factor, $v \equiv n/n_\phi$ ($n_\phi \equiv eB/h$ with Planck's constant $h$ and electric charge $e$). Thus, one can study how the SOC modifies $\Delta E_v$ at different $v$ and $B$ by measuring the thermal activation behavior of the minimum longitudinal resistance $R_{xx}^{min} \propto e^{-\Delta E_v/2kT}$. The study had indeed provided useful insights about the change of the graphene band in the presence of the SOC,[23] but it involves a large set of multiple measurements to obtain $R_{xx}$ as a function of $n$ at different $T$ and $B$.

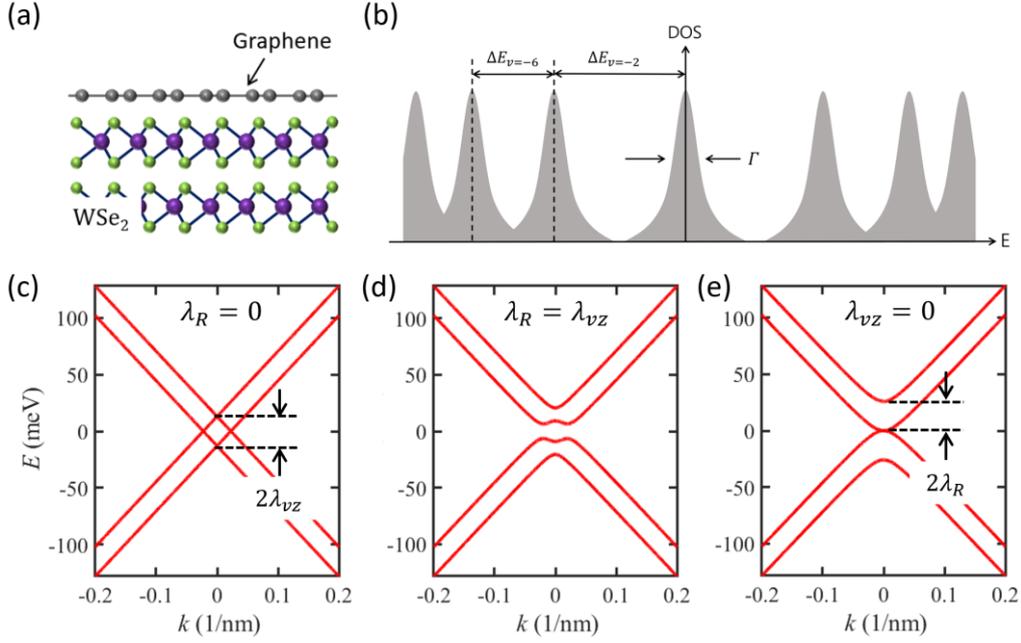

**Figure 1.** (a,b) Schematics of graphene-WSe$_2$ heterostructure (a) and electronic density of states (DOS) of an intrinsic graphene when Landau levels are formed (b). (c-e) Calculated band structure of monolayer graphene on TMDC at $K$ valley for $(\lambda_{vz}, \lambda_R) = (13.0, 0.0)$, $(9.2, 9.2)$, and $(0.0, 13.0)$ meV, respectively, at a fixed total SOC strength $\lambda = \sqrt{\lambda_{vz}^2 + \lambda_R^2} = 13$ meV in the absence of magnetic fields.

Here, we propose and demonstrate a simpler way to estimate the relative strengths of the SOC terms, $\lambda_{vz}$ and $\lambda_R$, in graphene-TMDC heterostructures from a Landau fan diagram $R_{xx}(n \text{ or } v, B)$ measured at a fixed $T$, without carrying out large sets of measurements. For this, we first use a simplified effective low-energy Hamiltonian[25,26] (**supplementary material**) to calculate the LL spectrums for different values of $\lambda_{vz}$ and $\lambda_R$, and show that, interestingly, the dispersions of the low-energy LLs below $|v| = 6$ sensitively depend on the SOC strengths (**Figure 2** and **3**). It indicates that one can use the Landau fan diagram to estimate their relative strengths. To demonstrate this, we further examined the calculated LL spectrums carefully and identified the four key features that are well separated from the complex background and thus can be compared directly with the experiment as listed below:

i) A gap opens at $v = 0$ more rapidly in $B$ as $\lambda_{vz}$ becomes larger, i.e., the graphene becomes insulating at lower $B$, unless $\lambda_{vz} \ll \lambda_R$;

ii) The QH signal—a dip in $R_{xx}$ and a plateau in the transverse resistance $R_{xy}$—appears at $v = \pm 3$ (and $\pm 5$) from low $B$ only when $\lambda_{vz}, \lambda_R \neq 0$;

iii) The QH signal at $v = \pm 4$ becomes stronger than the signal at $v = \pm 6$, i.e., it appears at lower $B$, only when $\lambda_R \approx 0$;

iv) As $\lambda_R$ increases, the QH signals at $v = \pm 4$ and $\pm 6$ will disappear and appear again at a larger and smaller $B$, respectively.

After identifying these key features, we examined a Landau fan diagram measured in one of the graphene-WSe$_2$ heterostructures used in our previous study[13] and found that $\lambda_R > \lambda_{vz}$, which is consistent with the results from other measurements done on the same sample previously.[13] We believe this simple strategy can be used to study how the relative strength of the Rashba and spin-valley Zeeman SOCs varies with different experimental conditions, such as twist angle,[27-29] pressure,[30] and strain,[31,32] in graphene-TMDC heterostructures quickly before carrying out more in-depth studies.

## 2. Theoretical Analysis
### 2.1 Band Structure of Monolayer Graphene on TMDC at Zero Magnetic Fields

To show that the Rashba and spin-valley Zeeman SOCs can affect the graphene band differently, we first calculate the band structure of graphene on TMDC at zero $B$ for different values of $\lambda_{vz}$ and $\lambda_R$ using the effective Hamiltonian,[25,26]

$$H = H_0 + \Delta\sigma_z + H_{KM} + H_{vz} + H_R, \quad (1)$$

where $H_0$ is the graphene's Dirac Hamiltonian and $\Delta\sigma_z$ is a symmetry breaking term that originates from the TMDC substrate which is nearly zero due to the large lattice mismatch between graphene and TMDC.[12] We have also ignored the intrinsic Kane-Mele SOC term, $H_{KM} = \lambda_{KM}\sigma_z\tau_z s_z$,[33] as it is known to be extremely small in graphene (only a few μeV).[34] The calculated graphene bands for $(\lambda_{vz}, \lambda_R) = (13.0, 0.0), (9.2, 9.2),$ and $(0.0, 13.0)$ meV are plotted in **Figure 1(c-e)**, respectively. The plots clearly show that while the spin-valley Zeeman SOC splits the band in energy by keeping its linear dispersion (**Figure 1(c)**), the Rashba term creates a nearly parabolic dispersion around the charge neutrality point (**Figure 1(e)**). Moreover, only when both SOC terms coexist, the gap at the charge neutrality point opens (**Figure 1(d)**). This is because the spin-valley Zeeman term $H_{vz}$ splits the spin up (↑) and down (↓) bands (so, it keeps the linear dispersion), whereas the Rashba term $H_R$ mixes the in-plane spin (i.e., the superposition of ↑ and ↓ spins) and sublattice degrees of freedom (*A* and *B*). Thus, without the broken sublattice symmetry term ($\Delta\sigma_z$) and the intrinsic Kane-Mele

SOC ($H_{KM}$), the gap can be opened only when both terms exist. This is consistent with previous calculations,[25] supporting the validity of our low-energy effective Hamiltonian given in Equation (1) and indicating that the Rashba and spin-valley Zeeman SOC modifies graphene band differently.

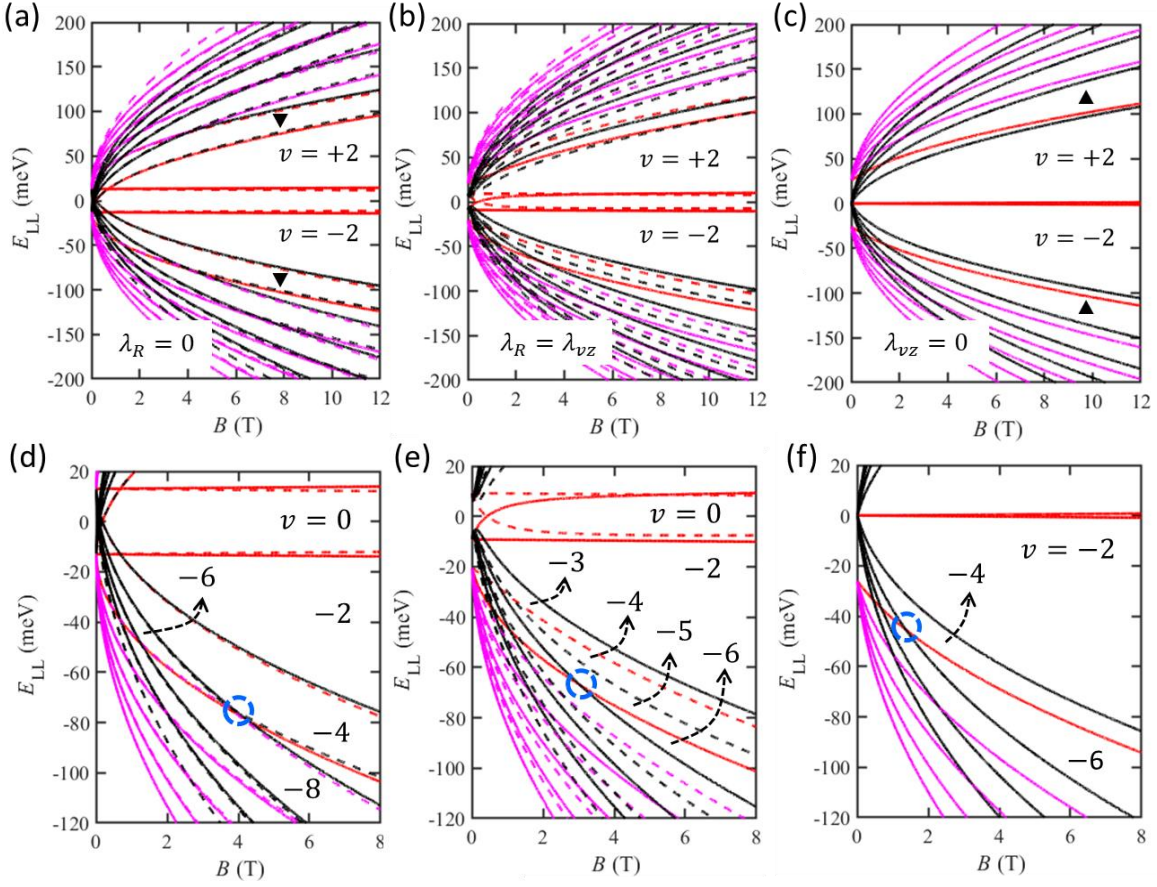

**Figure 2.** Effect of SOC on LL spectrum. (a-c) Calculated LL spectrums at positive $B$, for $(\lambda_{vz}, \lambda_R) = (13.0,0.0)$, $(9.2,9.2)$, and $(0.0,13.0)$ meV, corresponding to Figure 1(c-e), respectively. Red lines represent the $n = 0$ and $-1$ LLs, whereas the black and magenta lines are those for higher $n \geq 1$. Solid (broken) lines are the LLs from $K$ ($K'$) valley. Note that in (c), the solid and broken lines are perfectly merged together. The down- and up-triangles in (a) and (c) mark the LL energy gap at $v = \pm 4$ and $v = \pm 6$, respectively. (d-f) Zoomed-in view of LLs near zero energy shown in (a-c), respectively. Broken circles mark the position at which $\Delta E_{v=-6}$ becomes zero. See the text for more details.

**2.2 The Effects of SOC on the Landau Levels**

To investigate how these SOC terms affect the LL spectrum in the system, we followed the existing studies[26,31] to derive the effective Hamiltonian in the QH regime at finite $B$ (see **supplementary material** for details), and calculated the LL spectrums for $(\lambda_{vz}, \lambda_R) =$ (13.0,0.0), (9.2,9.2), and (0.0,13.0) meV as shown in **Figure 2(a-c)**, respectively. The results show rich features with clear differences between the three cases, suggesting that by studying the LL spectrum, one can estimate the SOC strengths in the system. For example, in all cases, $v = \pm 2$ states are the strongest with the largest energy gaps because the SOC strength used in the calculation, 13 meV, is still much smaller than the energy of the first excited LL in monolayer graphene (~36 meV at 1 T).[35,36] However, at larger $v$, i.e., when the SOC strength becomes comparable to the half of the $\Delta E_v$ of the intrinsic graphene, complex features appear. First, when only $\lambda_{vz}$ term exists (**Figure 2(a)**), the next filling factor at which the QH signal—minimum $R_{xx}$ and quantized $R_{xy}$—would appear is $v = \pm 4$ (marked by the down-triangles) not $v = \pm 6$ as $\Delta E_{v=\pm 4} > \Delta E_{v=\pm 6}$. On the other hand, when $\lambda_{vz} = 0$ (**Figure 2(c)**), $\Delta E_{v=\pm 6} > \Delta E_{v=\pm 4}$, so the $v = \pm 6$ state would appear (marked by the up-triangles) not the $= \pm 4$. When both terms coexist (**Figure 2(b)**), the energy gaps between the LLs are not clearly visible except for the one at $v = \pm 2$, leading to a complex LL spectrum at larger $|v|$. Such a sensitive dependence of $\Delta E_v$ on $\lambda_{vz}$ and $\lambda_R$ is indeed the main reason why the LL spectroscopy works.[23]

Interestingly, while examining the LL spectrum more closely (see **Figure 2(d-f)** for the zoomed-in view) and comparing the $\Delta E_v$ calculated for different sets of $\lambda_{vz}$ and $\lambda_R$ (**Figure 3**), we found that some low-energy LLs at $|v| \leq 6$ are separated from the complex background well enough to be compared with the experiment directly. First, as shown in **Figure 2(d-f)** and **3(a)**, when $\lambda_{vz}$ is finite (unless $\lambda_{vz} \ll \lambda_R$), a large gap opens at $v = 0$ from a low $B$ and increases more rapidly in $B$ as the $\lambda_{vz}$ gets larger (reaching $\Delta E_{v=0} \approx 26$ meV at ~1 T when $\lambda_{vz} = 13$ meV). On the other hand, when $\lambda_{vz} = 0$, only a small gap opens due to the Zeeman effect. This is because the spin-valley Zeeman SOC ($H_{vz} = \lambda_{vz} \tau_z s_z$) splits the linear graphene bands in energy (see **Figure 1(a)**) and thus creates two charge neutrality points from which the two sets of zero energy LLs will appear, leading to the large gap at zero density. Second, **Figure 2(d-f)** and **3(b)** show that the gap at $v = \pm 3$ and $\pm 5$, $\Delta E_{v=\pm 3, \pm 5}$, is finite at low $B$ only when both $\lambda_{vz}$ and $\lambda_R$ are finite, indicating that the QH signals would appear at $v = \pm 3$ and $\pm 5$ at low $B$ only when $\lambda_{vz}, \lambda_R \neq 0$. Third, **Figure 2(d-f)** and **3(c,d)** indicate that when $\lambda_R \approx 0$, the $\Delta E_{v=\pm 4}$ is larger than $\Delta E_{v=\pm 6}$, and it is suppressed as $\lambda_R$ increases. Thus, the QH signal at $v = \pm 4$ becomes stronger than the signal at $v = \pm 6$, i.e., it appears at lower $B$ when $\lambda_R \approx 0$. Lastly, we found that as $\lambda_R$ increases, $\Delta E_{v=\pm 4}$ and $\Delta E_{v=\pm 6}$ becomes zero at lager and smaller $B$, respectively (**Figure 3(c,d)**). This means that the QH signal at $v = \pm 4$ (and $\pm 6$) should exhibit a crossing (a dip in $R_{xx}$ disappears and appears again) at a higher (and lower) $B$ for a larger $\lambda_R$. We note that these features are well separated from other LLs and can be compared directly with the experiment without quantitatively detecting $\Delta E_\nu(B)$, as only the qualitative comparison is sufficient to estimate the relative strengths of the SOC terms.

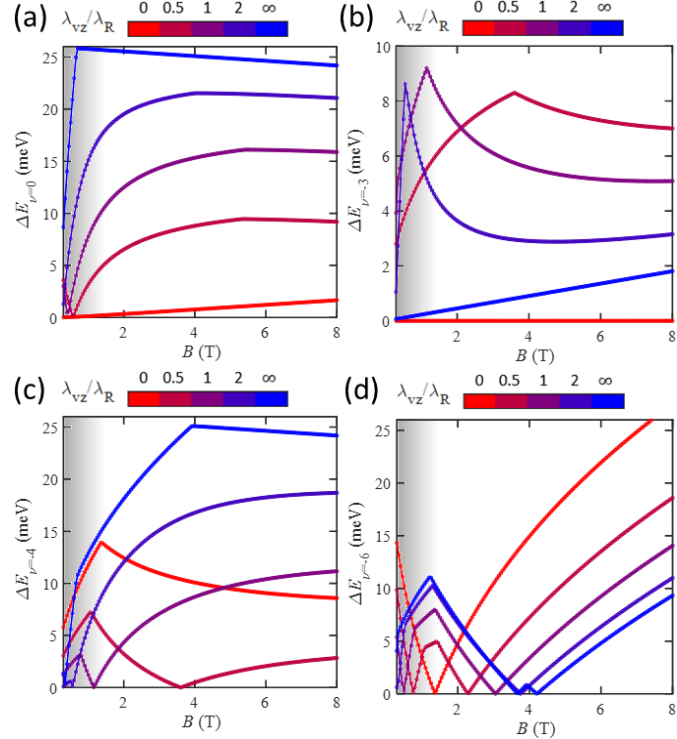

**Figure 3.** (a-d) Magnetic field dependence of the calculated LL gaps, $\Delta E_\nu(B)$, at $\nu = 0, -3, -4,$ and $-6$, respectively, for $\lambda_{vz}/\lambda_R = 0, 0.5, 1, 2,$ and $\infty$ (from blue to red) when $\lambda = 13$ meV. A shadowed area below $B \sim 1$ T represents a low magnetic field range in which the QH signals are weak (or do not appear) in general due to disorder.

In summary, after examining the corresponding energy scale, the robustness of the features and considering the electron-hole symmetry, we identify four features at low magnetic field range that are most practical in comparing with the experiments, i) the resistance at zero density, the QH signals at ii) $\nu = \pm 3$ (and $\pm 5$), iii) $\nu = \pm 4$, and iv) $\nu = \pm 6$. If we find a large resistance (or a clear insulating behavior) at zero density and a clearer QH signal at $\nu = \pm 4$ at lower $B$ than that at $\nu = \pm 6$, we can estimate $\lambda_{vz} > \lambda_R$, whereas observing a stronger signal at $\nu = \pm 6$ than the others indicates $\lambda_R \gg \lambda_{vz}$. Moreover, finding the QH signal at $\nu = \pm 3$ (and $\pm 5$) and the LL crossing at $\nu = \pm 4$ points to the coexistence of the Rashba and spin-valley Zeeman terms.

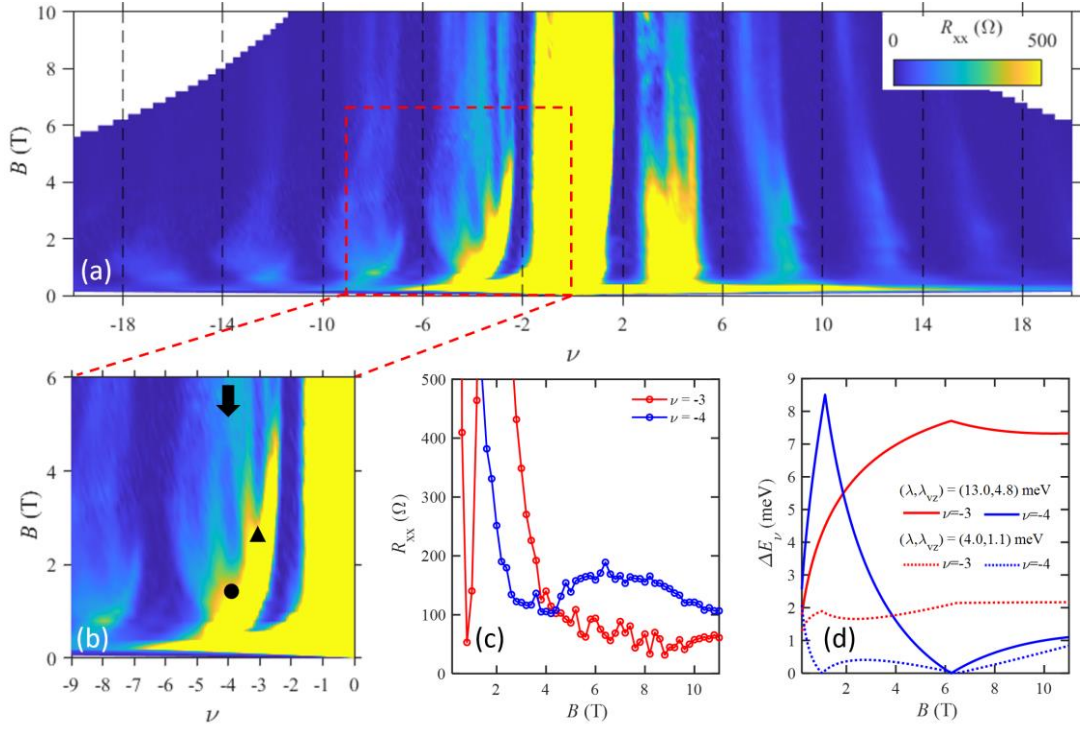

**Figure 4.** Comparison with experimental data. (a) A Landau fan diagram, a colormap of $\log(R_{xx})$ as a function of $B$ and $\nu$, and (b) its zoomed-in view in an area enclosed by the red broken box. The data is taken from our previous study[13] (see **supplementary material** for details). Filled up-triangle, circle, and diamond mark the $=-3, -4$, and $-5$ QH signals, respectively. An arrow indicates the suppression of $\nu = -4$ LL above ~4 T. (c) 1D cuts of the data shown in (a) at $\nu = -3$ and $-4$. (d) Magnetic field dependence of $\Delta E_{\nu=-3}$ (red) and $\Delta E_{\nu=-4}$ (blue) for $(\lambda, \lambda_{vz}) = (13.0, 4.8)$ and $(4.0, 1.1)$ meV in solid and broken lines, respectively.

## 3. Experimental Signature and Comparison

Having identified the four key features, we turn our attention to the Landau fan diagram, $R_{xx}(\nu, B)$, measured in one of our graphene-WSe$_2$ heterostructures at 1.5 K (see **supplementary material** for experimental details).[13] **Figure 4(a)** plots the result from which we can clearly identify the dark vertical stripes at $\nu = \pm 2, \pm 6, \pm 10, \pm 14, ...$ that matches with the QH sequence of the pristine monolayer graphene and find that the resistance peak at $\nu = 0$ remains small and it does not increase in $B$ (the criteria i). Moreover, we find that the QH signal at $\nu = \pm 6$ appears with a smaller $R_{xx}^{min}$ at lower $B$ than the signal at $\nu = \pm 4$ (see **Figure 4(b)** for zoomed-in view; criteria iii). These two findings indicate that our sample has a stronger Rashba SOC, i.e., $\lambda_R \gg \lambda_{vz}$. However, between $\nu = -2$ and $-6$ (**Figure 4(b)**), we can also identify the QH signals at $\nu = -3$ (up-triangle), $-4$ (circle), and $-5$ (diamond), and additionally, find that the $\nu = -4$ state disappears upon increasing $B$ above ~4 T due to the LL crossing (the criteria ii and iv). These findings indicate that although $\lambda_{vz}$ is much smaller than $\lambda_R$, it is finite.

For more detailed analysis, we plot the 1D cuts of $R_{xx}(B)$ along $v = -3$ and $-4$ in **Figure 4(c)**. The plot clearly shows that the $R_{xx}^{min}$ at $v = -4$ peaks at around 6 T, indicating the LL crossing while the $R_{xx}^{min}$ at $v = -3$ decreases continuously as $B$ increases. By varying both $\lambda_{vz}$ and $\lambda_R$, we find that when $(\lambda, \lambda_{vz}) = (13.0, 4.8)$ in meV, the gap at $v = -4$, $\Delta E_{v=-4}$, becomes zero around 6 T and $\Delta E_{v=-3}$ keeps increasing in $B$ that matches well with the observed behavior (**Figure 4(d)**). This is indeed consistent with the results from the ballistic transport spectroscopy measurements done on the same sample previously.[13] However, it is worth noting that to determine the absolute values of the SOC strengths, we need to measure the thermal activation energy of the LLs at each $v$, $\Delta E_v$, study how it varies with $B$, and compare it with the theoretical curves shown in **Figure 3**. This is because as long as $\Delta E_v$ becomes larger than the thermal energy or disorder level (i.e., $\Delta E_v > \max(kT, \Gamma)$), the QH signal will appear at the corresponding $v$, so one cannot trace how the $\Delta E_v$ varies in $B$ quantitatively. For instance, the broken lines in **Figure 4(d)** plot the $\Delta E_{v=-3}$ and $\Delta E_{v=-4}$ for $(\lambda, \lambda_{vz}) = (4.0, 1.1)$ meV that show a similar behavior as those calculated for $(\lambda, \lambda_{vz}) = (13.0, 4.8)$ meV but at a much smaller energy scale. Nonetheless, we have clearly demonstrated that by checking the four key features in a single Landau fan diagram measured at a fixed $T$ (**Figure 4(a)**), one can directly estimate the relative strength of the Rashba and spin-valley Zeeman SOCs. This can be particularly useful when studying how the SOC strengths vary with external conditions, such as twist angle,[27-29] pressure,[30] or strain,[31,32] as it is not practical to measure the thermal activation gap ($\Delta E_v$) for all different parameters. One can therefore use the four criteria proposed here to choose few parameters that may exhibit clear differences in SOC strengths to carry out more in-depth measurements.

## 4. Conclusion

In conclusion, by carefully analyzing the LL spectrum of graphene-TMDC heterostructures, we demonstrate a practical way to compare the experimental data with calculations for the LL spectroscopy of spin-orbit-coupled bands in graphene. From the theoretical analysis, we determined four robust features, i) the resistance at zero density, the QH signals at ii) $v = \pm 3$ (and $\pm 5$), iii) $v = \pm 4$, and iv) $v = \pm 6$, that can be used to identify the relative strengths of the Rashba and spin-valley Zeeman SOC terms in the system (**Figures 1-3**). In the experiment (**Figure 4**), we found a LL sequence at $v = \pm 2, \pm 6, \pm 10, \pm 14, ...$ that resembles those of monolayer graphene and no clear gap opening at zero density, suggesting a stronger Rashba SOC. Moreover, we found that the $v = -4$ state appears at low $B \sim 2$ T and disappears around 6 T and that the $v = -3$ and $-5$ states start to appear from a small $B$. This indicates that although the spin-valley Zeeman SOC is much weaker than the Rashba term, it exists in our sample, consistent with the results from the spectroscopic measurements done on the same sample.[13] Although more in-depth studies are needed to estimate the exact values of the SOC strengths, we believe our work provides a simple and practical way to analyze the QH data in graphene-TMDC heterostructures that can be used to investigate how the SOC strength varies with external parameters, such as twist angle,[27-29] pressure,[30] or strain,[31,32] more conveniently.

**Acknowledgements**

The work is financially supported by the University Grants Committee/Research Grant Council of Hong Kong SAR under schemes of Area of Excellence (AoE/P-701/20), ECS (27300819), and GRF (17300020, 17300521, 17309722). We also acknowledge financial support from University Research Committee (URC) of The University of Hong Kong under the schemes of Seed Fund for Basic Research (202111159043) and Seed Funding for Collaborative Research.

# Supplementary Material

## 1. Landau Levels of Graphene with Proximitized Spin-orbit Coupling (SOC)

To calculate the Landau levels, we follow the existing studies[26,31] where bosonic Landau ladder operators $a$ and $a^\dagger$ were derived. For a positive $B$, they are defined as

$$\hbar v_F \left(k_x + \frac{eBy}{\hbar} + ik_y\right) \to \hbar \omega_B a, \quad \hbar v_F \left(k_x + \frac{eBy}{\hbar} - ik_y\right) \to \hbar \omega_B a^\dagger,$$

with a cyclotron energy $\hbar \omega_B = \sqrt{2\hbar eB} v_F$ and $\hbar v_F = \sqrt{3}at/2$ ($a \approx 0.246$ nm: lattice constant of graphene, $t \approx 2.7$ eV: nearest-neighbor hopping strength). Since these ladder operators are the same as those in harmonic oscillator, one can choose the eigenfunctions $|n\rangle$ ($n$: a non-negative integer) that satisfies $a|n\rangle = \sqrt{n}|n-1\rangle$ and $a^\dagger|n\rangle = \sqrt{n+1}|n+1\rangle$ and use a spinor ansatz to set the basis of Hamiltonian at finite $B$ as[26]

$$\psi^K_{B>0,n,m} = \begin{pmatrix} C^{A\uparrow}_{n,m}|n\rangle \\ C^{A\downarrow}_{n,m}|n+1\rangle \\ C^{B\uparrow}_{n,m}|n-1\rangle \\ C^{B\downarrow}_{n,m}|n\rangle \end{pmatrix}, \quad \psi^{K'}_{B>0,n,m} = \begin{pmatrix} C^{A\uparrow}_{n,m}|n-1\rangle \\ C^{A\downarrow}_{n,m}|n\rangle \\ C^{B\uparrow}_{n,m}|n\rangle \\ C^{B\downarrow}_{n,m}|n+1\rangle \end{pmatrix}, \quad (S1)$$

where $m$ is a subset of solutions for orbital index $n$, one can derive effective low-energy Hamiltonians in the quantum Hall regime for both $K$ and $K'$ valleys as follows

$$H^K_{B>0,n} = \begin{pmatrix} \lambda_{vz} + E_z & 0 & \hbar\omega_B\sqrt{n} & 2i\lambda_R \\ 0 & -\lambda_{vz} - E_z & 0 & \hbar\omega_B\sqrt{n+1} \\ \hbar\omega_B\sqrt{n} & 0 & \lambda_{vz} + E_z & 0 \\ -2i\lambda_R & \hbar\omega_B\sqrt{n+1} & 0 & -\lambda_{vz} - E_z \end{pmatrix}, \quad (S2)$$

$$H^{K'}_{B>0,n} = \begin{pmatrix} -\lambda_{vz} + E_z & 0 & -\hbar\omega_B\sqrt{n} & 0 \\ 0 & \lambda_{vz} - E_z & 2i\lambda_R & -\hbar\omega_B\sqrt{n+1} \\ -\hbar\omega_B\sqrt{n} & -2i\lambda_R & -\lambda_{vz} + E_z & 0 \\ 0 & -\hbar\omega_B\sqrt{n+1} & 0 & \lambda_{vz} - E_z \end{pmatrix}, \quad (S3)$$

where $E_z = g\mu_B B \approx 0.113 \times B$(in T) meV is the Zeeman energy with an electron spin factor $g \approx 1.95$[37] and Bohr magneton $\mu_B = e\hbar/2m_e$. The Hamiltonians above can be solved numerically for $n \geq 1$ to get LL spectrum for a given set of $(\lambda_{vz}, \lambda_R)$.

Interestingly, after carefully examining the Equation (S1-3) at low energy ($n \leq 0$), we find that there are non-trivial solutions with interesting properties. First, for $n = -1$, the wavefunctions

$$\psi^K_{B>0,n=-1,m} = \begin{pmatrix} 0 \\ |0\rangle \\ 0 \\ 0 \end{pmatrix}, \quad \psi^{K'}_{B>0,n=-1,m} = \begin{pmatrix} 0 \\ 0 \\ 0 \\ |0\rangle \end{pmatrix}$$

that have only $A \downarrow$ component at $K$ valley and $B \downarrow$ component at $K'$ valley satisfy the eigenvalue equation for $H^K_{B>0,n}$ and $H^{K'}_{B>0,n}$ with eigenvalues

$$E^K_{B>0,n=-1} = -\lambda_{vz} - E_z, \qquad E^{K'}_{B>0,n=-1} = \lambda_{vz} - E_z$$

that vary in $B$ only by Zeeman energy. Furthermore, at $n = 0$, the Hamiltonians $H^K_{B>0,n}$ and $H^{K'}_{B>0,n}$ become

$$H^K_{B>0,n=0} = \begin{pmatrix} \lambda_{vz} + E_z & 0 & 0 & 2i\lambda_R \\ 0 & -\lambda_{vz} - E_z & 0 & \hbar\omega_B \\ 0 & 0 & \lambda_{vz} + E_z & 0 \\ -2i\lambda_R & \hbar\omega_B & 0 & -\lambda_{vz} - E_z \end{pmatrix},$$

$$H^{K'}_{B>0,n=0} = \begin{pmatrix} -\lambda_{vz} + E_z & 0 & 0 & 0 \\ 0 & \lambda_{vz} - E_z & 2i\lambda_R & -\hbar\omega_B \\ 0 & -2i\lambda_R & -\lambda_{vz} + E_z & 0 \\ 0 & -\hbar\omega_B & 0 & \lambda_{vz} - E_z \end{pmatrix}$$

that can be block-diagonalized into $1 \times 1$ and $3 \times 3$ matrices. From the $1 \times 1$ matrices, we get additional two non-trivial eigenvalues for $B \uparrow$ state at $K$ valley and $A \uparrow$ state at $K'$ valley that depends only on $E_z$ and $\lambda_{vz}$,

$$E^K_{B>0,n=0} = \lambda_{vz} + E_z, \qquad E^{K'}_{B>0,n=0} = -\lambda_{vz} + E_z.$$

The remaining $3 \times 3$ matrices can be solved numerically. The calculated LL spectrums are shown in **Figure 2**.

## 2. Device Fabrication and Measurement

The graphene-WSe$_2$ device was fabricated using conventional nanofabrication technique.[13] We first exfoliated hBN, WSe$_2$, and graphene flakes onto silicon wafers, and after examining them under an optical microscope, target flakes were stacked into hBN-graphene-WSe$_2$-hBN heterostructure on a highly doped silicon substrate with 285-nm-thick oxide following the standard van der Waals (vdW) assembly method.[38] Thermal annealing was done at 250 °C for 2 h in a tube furnace in Ar/H2 forming gas to reduce the disorders in the stack. After the annealing, we fabricate 1D electrical contacts using the standard electron-beam lithography, reactive-ion etching (CF4/O$_2$ mixture gas with flow rates of 5/25 sccm, RF power: 60 W) and electron beam evaporation (5 nm Cr and 50 nm Au films).[38] Finally, we shape the device into a Hall bar geometry through a second electron-beam lithography and reactive-ion etching.

For electrical measurement, we put the device in a 1.5 K cryogen-free variable temperature insert (VTI) with a 14-T superconducting magnet. The longitudinal resistance $R_{xx} = V_{xx}/I_{xx}$ was measured by applying a small low-frequency (17.777 Hz) AC voltage of 500 μV between the source and drain contacts while simultaneously measuring the current $I_{xx}$ in the circuit and the voltage drop $V_{xx}$ between another two Hall probes using two lock-in amplifiers (Stanford Research SR830). Meanwhile, to obtain the Landau fan diagram $R_{xx}(v, B)$, we sweep the back gate and change the magnetic field in steps by controlling the Keithley 2400 source-meter and the superconducting magnet, respectively. The device image and measurement configuration are shown in **Figure S1**.

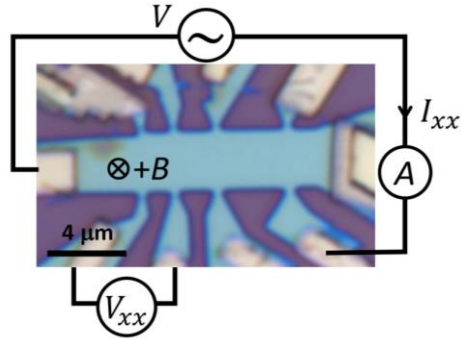

**Figure S1**. An optical microscope image of the device with a configuration to measure longitudinal resistance $R_{xx} = V_{xx}/I_{xx}$ under perpendicular magnetic fields $B$. The charge density ($n$) is controlled by applying a voltage to the silicon back gate (not shown).